\documentclass[aip,aps,prb,twocolumn,a4paper,10pt,showkeys,showpacs,preprintnumbers,floatfix,superscriptaddress]{revtex4-1}

\usepackage{graphicx}		% Include figure files
\usepackage{dcolumn}		% Align table columns on decimal point
\usepackage{bm}				% bold math
\usepackage{amsmath}
\usepackage{amssymb}
\usepackage{amsfonts}
\usepackage{times}
\usepackage{tabularx}
\usepackage{xcolor,colortbl}
\definecolor{lightgray}{gray}{0.9}

\begin{document}

\title{Dynamical influence of vortex-antivortex pairs in magnetic vortex oscillators}

\author{R. M. Otxoa}
\email{ro274@cam.ac.uk}
\affiliation{Institut d'Electronique Fondamentale, UMR CNRS 8622, Univ. Paris-Sud, 91405 Orsay, France}
\affiliation{Hitachi Cambridge Laboratory, J. J. Thomson Avenue, CB3 OHE, Cambridge, United Kingdom}

\author{S. Petit-Watelot}
\affiliation{Institut d'Electronique Fondamentale, UMR CNRS 8622, Univ. Paris-Sud, 91405 Orsay, France}
\affiliation{Institut Jean Lamour, UMR CNRS 7198, Univ. de Lorraine, F-54506 Vandoeuvre-les-Nancy, France}

\author{M. Manfrini}
\author{A. Thean}
\author{I. P. Radu}
\affiliation{IMEC, Kapeldreef 75, 3001 Leuven, Belgium}

\author{Joo-Von Kim}
\author{T. Devolder}
\affiliation{Institut d'Electronique Fondamentale, UMR CNRS 8622, Univ. Paris-Sud, 91405 Orsay, France}

\date{\today}

\begin{abstract}

We study the magnetization dynamics in a nanocontact magnetic vortex oscillators as function of temperature. Low temperature experiments reveal that the dynamics at low and high currents differ qualitatively. At low currents, we excite a temperature independent standard oscillation mode, consisting of a gyrotropic motion of a vortex about the nanocontact in the free layer. Above a critical current, a sudden jump in the frequency is observed, which occurs with a substantial increase of the frequency versus current slope factor. Using micromagnetic simulation and analytical modeling, we associate this new regime to the creation of a vortex-antivortex pair in the pinned layer of the spin valve. This pair gives an additional  perpendicular spin torque component that alters the free layer vortex dynamics, which can be quantitatively accounted for by an analytical model.
\end{abstract}

\maketitle

\section{Introduction}

Magnetic vortices are fundamental topological states in restricted geometries such as thin submicron dots or nanopillars.\cite{COWBURN, Pribiag} For a certain range of aspect ratios, the micromagnetic ground state is a vortex structure because the circular configuration of the spins minimizes the stray dipolar fields. Because the norm of the local magnetization vector is conserved in strong ferromagnets due to a large exchange interaction, the magnetization at the core of the vortex tilts out of the film plane.

Vortices can also be nucleated in extended ferromagnetic thin films in the nanocontact (NC) geometry \cite{Pufall:PRB:2007, Mistral:PRL:2008}, in which it has been demonstrated that vortex manipulation over micrometer-scale distances is possible~\cite{ManNat}. Spin torque effects appear when large current densities ($\sim 10^{12}$ A/m$^2$) are applied through the NC, which also results in significant Oersted-Amp\`{e}re fields, e.g., 300 mT for 50 mA in 100 nm diameter NCs. Indeed the perpendicular component of this current flow (relative to the film plane) leads to a circulating Oersted-Amp\`{e}re field akin to that generated by a cylindrical conductor. As such, a vortex state can appear by minimizing the Zeeman energy associated with the Oersted fields, in contrast to the case of confined geometries in which stray fields are minimized. Because the Zeeman interaction is proportional to the current, vortices only appear above a certain threshold or nucleation current ($I_{\text{nucl}}$)~\cite{Mistral:PRL:2008}.

However, processes involving vortex nucleation are subject to conservation laws involving topological charges \cite{Devolder:APL:2009}. The topology involved is described by the Skyrmion number, $q = \eta p/2$, where $\eta$ is the vorticity which describes the curling magnetization of a vortex by $\eta$ = +1 and $\eta$ = -1 for an antivortex. $p$ is the core polarity, which describes the orientation of the magnetization at the vortex core. Since a uniform state has a total Skyrmion number (topological charge) of $q = 0$, the nucleation of a vortex ($q = p/2$) must be accompanied by the nucleation of an antivortex with the same core polarity such that the total $q$ remains zero~\cite{Tretiakov:PRB:2007}.

The stability of a vortex-antivortex (V-AV) pair in thin films strongly depends on the boundary conditions, i.e., on the micromagnetic state at the system edges. It has been shown \cite{Devolder2010} that pair nucleation in the magnetic free layer in zero field is followed by the antivortex being expelled by the Oersted-Amp\`{e}re field, resulting in steady-state oscillations of the vortex around the NC~\cite{Kim:PRB:2010}. The resulting micromagnetic state of the free layer therefore resembles the well-known Landau state. In contrast, the presence of an in-plane magnetic field would favor an uniform magnetic state far from the NC. In this case, the antivortex would be bound to the vortex such that the uniform state is preserved in the bulk of the film. This should equally be true for a ferromagnet exchange-biased by an antiferromagnet, where the internal field acting on the ferromagnet due to the exchange coupling also favors a uniform state.

Previous studies \cite{RotoloTM,Kuepferling} have suggested that the two ferromagnetic layers of a nanocontacted spin-valve structure may contain a vortex state under certain conditions of the applied magnetic field and injected current. To date, a successful model to explain the new observed dynamics has been lacking. In this article, we address this question from experimental and theoretical perspectives. We present experimental evidence of vortex-antivortex pair nucleation in the pinned layer of magnetic nanocontacts. At low bias currents, we observe the usual free-layer vortex oscillations expected of such structures~\cite{RubenPSB}. However, above a certain critical current, we detect the presence of a vortex-antivortex pair in the pinned layer through changes in the power spectra associated with the free-layer vortex oscillations. This critical current is strongly temperature-dependent, which is suggestive of a thermally-activated process. The experimental results are consistent with predictions based on rigid-vortex model.

\section{Experimental Results}

Figure 1 shows the experimental system studied. 
\begin{figure}%[htb]
\includegraphics[width=7.5cm]{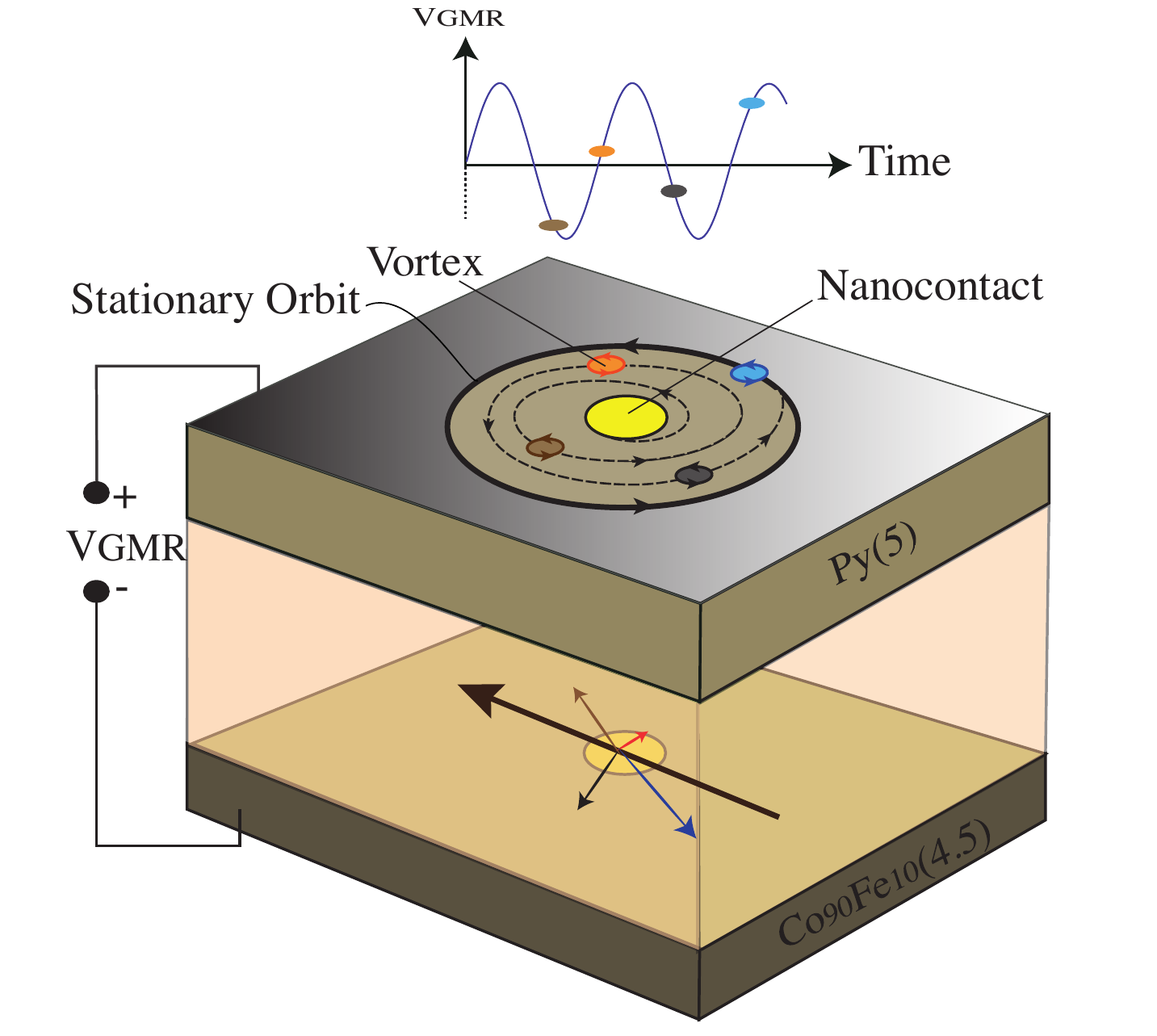}
\caption{(Color online) Vortex oscillations in the free layer of the magnetic nanocontact device, with uniform magnetization in the pinned CoFe layer. The vortex spirals out until a stationary orbit in a potential provided by the Oersted-Amp\`{e}re field. The vortex orbital motion leads to a time-varying voltage (top inset). Arrows represent the projection of the average free layer magnetization underneath the NC area.}
\label{standardmode}
\end{figure}
It consists of a metallic nanocontact fabricated on top of a spin-valve (SV) stack of width $L=17 \mu$m. The composition of the SV is IrMn(6)/Co$_{90}$Fe$_{10}$(4.5)/Cu(3.5)/Ni$_{80}$Fe$_{20}$(5)/Pt(3), where the number in parentheses denote the layer thickness in nanometers. Details of the fabrication process are given elsewhere~\cite{Manfrini:APL:2009}. For the system studied here, the NC radius r$_{\text{nc}}$ is 75 nm~\cite{RubenPSB}. To determine the free layer magnetization (Ni$_{80}$Fe$_{20}$) and the Gilbert damping, ferromagnetic resonance experiments were performed from which we determined $\mu_{0}$M$_{\text{s}}$=1.1 T and $\alpha$=0.013, respectively. The electrical properties of the device were characterized by performing static magneto-transport measurements, which give a device resistance of 8.7 $\Omega$ and a magnetoresistance of 25 m$\Omega$ at room temperature. The Co$_{90}$Fe$_{10}$ layer acts as pinned layer since it is exchanged bias by an antiferromagnet (IrMn). Differential magnetoresistance curves have been also measured with a lock-in technique with a 10 $\mu$A ac current and zero dc applied current for different applied temperature. This technique allow us to find out how the exchange bias field ($\vec{H}_{\text{bias}}$) varies from 80 K up to 420 K, as it is shown in figure \ref{bias and Icrit}.

Prior to electrical characterization of the nanocontact device, we applied a magnetic field of $\approx$ 115 mT along the easy axis in order to saturate the free layer magnetization and therefore avoid to have any domain wall or vortex structure in the initial state. The device was measured at different temperatures from 6 K to 300 K in a cryostat probe station under zero applied field. To characterize the magnetization dynamics, the corresponding high-frequency fluctuations in the giant magnetoresistance signal were measured with a spectrum analyzer after amplification. The dc current ($I_{\textrm{dc}}$) applied to the nanocontact was ramped from 0 to 40 mA (upward scan) then back to 0 (downward scan), with electrons always flowing from the free to the pinned layer. The upward scans are used to determine the nucleation current $I_{\textrm{nucl}}$ of the free layer vortex, at which the voltage spectra exhibit a series of well-defined peaks representing the vortex gyration around the nanocontact~\cite{Devolder:APL:2009}.

\begin{figure}%[htb]
\includegraphics[width=9cm]{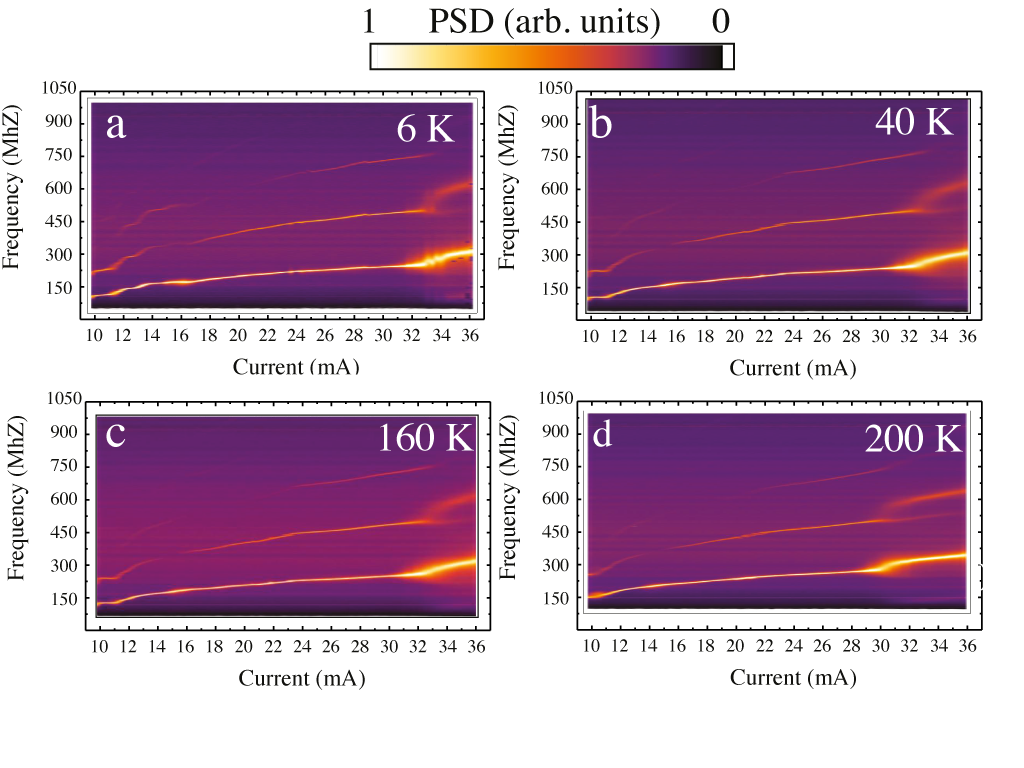}
\caption{\label{PSDtemp} Voltage power spectral densities (PSD) as function of current for different temperatures: (a) 6 K, (b) 40 K, (c) 160 K, and (d) 200 K.}
\end{figure}

\begin{figure}%[htb]
\includegraphics[width=9cm]{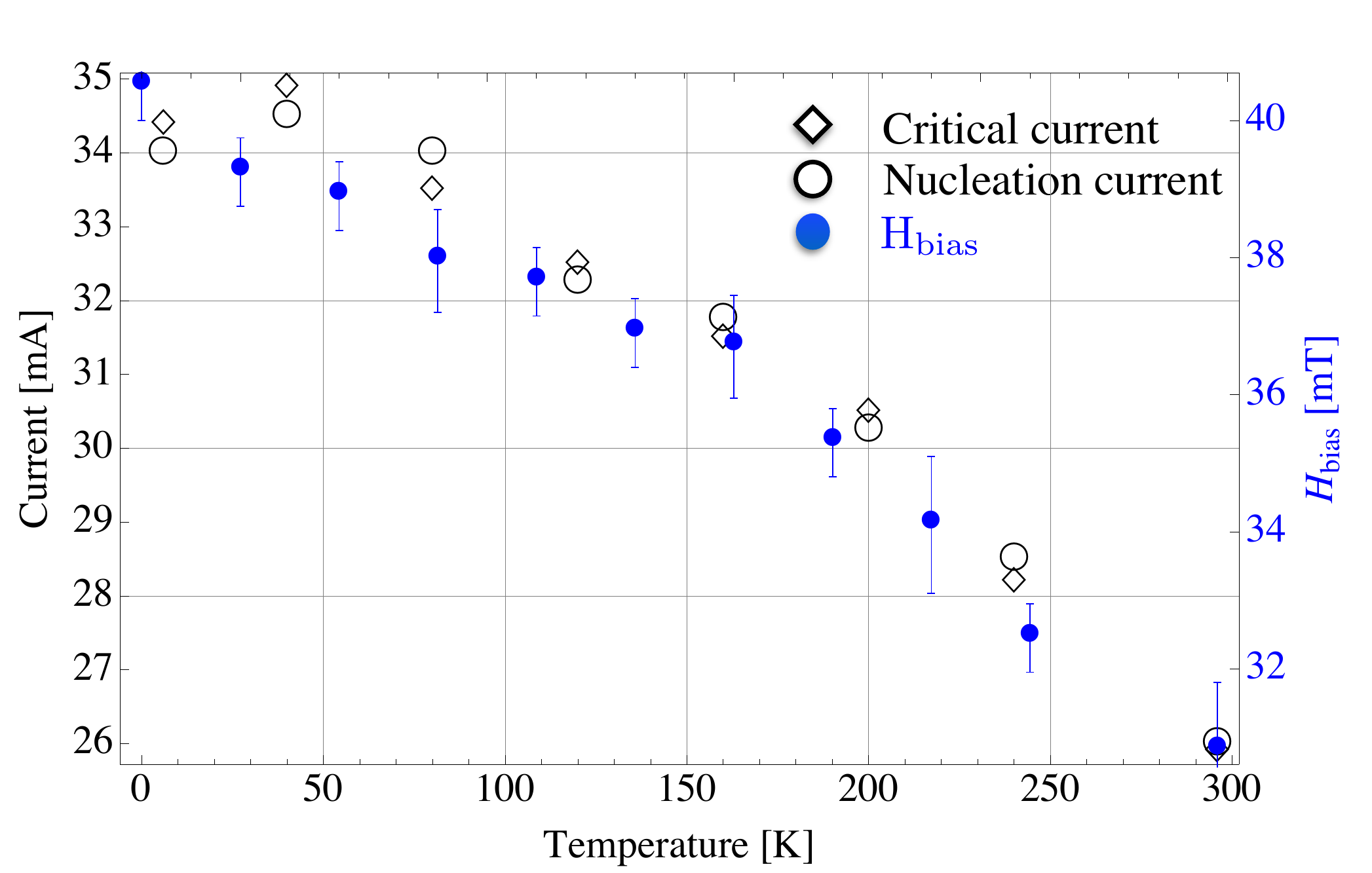}
\caption{\label{bias and Icrit} Critical current, nucleation current, and exchange bias field as a function of temperature.}
\end{figure}

Figure \ref{PSDtemp} shows the voltage power spectral density (PSD) as a function of the applied current, associated with oscillations of the vortex for four different temperatures, 6, 40, 160, and 200 K.
 The spectra are measured from the highest current value ($\approx$ 40 mA) to the annihilation current ($\approx$ 9 mA) of the vortex. Except for currents close to the annihilation of the vortex, where the dynamics is not well understood, a quasi-linear dependence of the oscillating frequency on current is observed. This result is consistent with a confining potential for the vortex dynamics that is determined by the Zeeman energy associated with the  Oersted-Amp\`{e}re field, in line with previous observations on other spin-valve compositions~\cite{Devolder:APL:2010,Manfrini:APL:2009}. This "standard" oscillation mode corresponds to a vortex that orbits around the nanocontact in the free layer (Fig. 1).

However, a different dynamical behavior is seen for currents above $I_{\text{dc}}$ $\approx$ 30 mA at different temperatures.  At this critical current (I$_{\text{crit}}$), and in the range of a few mA above this value, power spectra of the vortex oscillations exhibit a bimodal character. We interpret this as a signature of thermally-driven hopping between two distinct dynamical modes with different frequencies, which we label hereafter as the upper mode (UM) and lower mode (LM). Above this critical current, the LM frequency branch represents a continuation of the standard mode, while the UM branch occurs at a higher frequency with a different slope compared with the LM mode. While we have not obtained time-resolved measurements that confirm the hopping between the UM and LM, a signature of the hopping is present in the PSD shown in Figs 2(a)-(d). If a vortex is hopping between two oscillation modes, UM and LM, it spends some time $t_{\text{UM}}$ and $t_{\text{LM}}$ in each mode. A clear feature of hopping is a signal that should appear at frequencies related with the inverse time that the vortex spends to go from one mode to the other one. The observed frequency at which this mode appears is $\approx$ 50 MHz.  At high current there appears a low frequency shoulder that only survives as long as the UM and LM modes are present. We have gathered in Table I the frequency jumps $f_{\text{UP}}$-$f_{\text{LM}}$ at $I_{\text{crit}}$ and their slopes ($df/dI$) at different experimental temperatures.
The dependence of $I_{\text{nucl}}$ and $I_{\text{crit}}$ as function of the temperature is shown in Fig.~\ref{bias and Icrit}. It is worth noticing these two characteristic currents follow the same trend, which suggests they involve similar physical processes. While $I_{\text{nucl}}$ represents the threshold current to nucleate a vortex state in the free layer, the value $I_{\text{crit}}$ represents a threshold current associated with some dynamics in the pinned layer (PL). The observed decrease in power (not shown) when the current is increased is in agreement with this assertion.

The exchange bias field $\vec{H} _{\text{bias}}$ acting on the PL decreases as the temperature is increased. Note that the cryostat temperature in our experiment does not take into account the Joule heating resulting from the current flow through the nanocontact device. Recently, it has been reported how the current density is distributed along the SV stack after passing trough the NC \cite{APL:Sebastien:2012}. This study allowed us to simulate the temperature profile across the SV thickness underneath the NC \cite{SEbTEMP}. For currents of about 50 mA, we found the temperature increase to be between 150 K and 200 K in the vicinity of the NC. For example, at room temperature the $\vec{H}_{\text{bias}}$ is about 30 mT whereas for an injected current of 50 mA it would decreases down to 22 mT. Therefore, one consequence of Joule heating is to reduce the bias field acting on the PL in the nanocontact region. In line with this reasoning, nucleation of some magnetic structure with an out-of-plane magnetization may occur in the PL underneath the NC. This is discussed in more detail in Section IV.

The presence of a V-AV pair in the pinned layer would give rise to an out-of-plane component of magnetization in the nanocontact region. This has two consequences on the vortex dynamics. First, it leads to a coupling between the gyrating vortex in the free layer and the vortex-antivortex pair through the dipolar interaction, which could lead to an additional term in the confining potential for the vortex. Second, and more importantly, the core magnetization of the vortex-antivortex pair leads to a perpendicular-to-plane component for the spin-polarized current flowing between the free and pinned layers, which can generate additional spin torques for the vortex dynamics. The latter may explain the frequency jump at the threshold $I_{\text{crit}}$ along with the different slope $df/dI$ observed for the UM. In this light, the existence of two oscillation modes suggests that the vortex-antivortex pair in the pinned layer has a finite lifetime, where the hopping is due to the repeated nucleation and annihilation of the vortex-antivortex pair due to thermal fluctuations.

%%%%%%%%%%%%%%%%%%%%%%%%%%%%%%%%%%%
%Explanation of the model%
%%%%%%%%%%%%%%%%%%%%%%%%%%%%%%%%%%%
\section{Model for Vortex-Antivortex nucleation}

In order to quantify the scenario involving nucleation and annihilation of the vortex-antivortex pair described in the previous section, we have extended the Thiele formalism for describing the nanocontact vortex oscillations by accounting for the presence of the vortex-antivortex pair. We first examine the nucleation problem without and with spin transfer torques, and then provide a description of the vortex dynamics in the presence of the vortex-antivortex pair.

\subsection{Vortex-antivortex nucleation in the pinned layer in the absence of STT}

As shown previously~\cite{Devolder2010}, the onset of vortex oscillations in the free layer takes place after the following sequence: (1) the initial uniformly-magnetized state is distorted by the large Oersted-Amp{\`e}re field; (2) nucleation of a vortex-antivortex pair occurs after this distortion becomes irreversible; (3) expulsion of the antivortex away from the nanocontact region. A similar process is expected for vortex-antivortex pair nucleation in the pinned layer, but the key difference is the presence of the exchange bias field $\vec{H}_{\text{bias}}$ that will limit the separation distance between the vortex and the antivortex. The equilibrium separation will therefore be determined by balancing the competing attractive and repulsive forces.

% have shown the sequence that is followed to finally reach a vortex state. This sequence, is mediated by distortion of the free layer, the nucleation of a vortex-antivortex pair and finally the expulsion of the antivortex far away from the NC area. Since we are interested in the micromagnetic configuration of the pinned layer it is necessary to consider the important role of the exchange bias field $\vec{H}_{\text{bias}}$ in the whole process. The process of nucleation is interpreted as the separation of the vortex and the antivortex and therefore, there are competing forces that can be divided into attractive and repulsive.

%Once the V-AV pair is nucleated, the exchange energy E$_{\text{exc}}$ (short range attractive force) tends to keep the vortex and the antivortex closer. On the other hand the Zeeman energy coming from the Oersted field (intermediate range repulsive force) tries to separate them in order to keep the same symmetry of the Oersted field below the NC. The exchange bias field energy E$_{\text{eb}}$ (long range attractive force) tend to reduce the vortex-antivortex core distance because they are separated by a zone whose magnetization is opposite to the $\vec{H}_{\text{bias}}$.

To describe the relevant energies associated with nucleation, we use the rigid vortex model to describe the vortex-antivortex pair. This formalism allows us to express the relevant energies in terms of the positions of the vortex and antivortex cores by using a suitable \emph{ansatz} for the spin structure at the cores. To simplify the integrals for the energies, we assume that the vortex is centered about the nanocontact, $X_{v}$= $(0,0)$, while the antivortex is situated at $\vec{X}$=$(X,Y)$. The spatial distribution of the core magnetization in polar coordinates ($\Theta$ and $\Phi$) is taken to be,
\begin{equation}
\label{eq1}
\Phi(x,y;X,Y)=\eta \tan^{-1} \left( \frac{y-Y}{x-X}\right)+\frac{\pi}{2},
\end{equation}
where $\eta = \pm 1$ is the vorticity. By using this \emph{ansatz}, the total magnetic energy of the pinned layer,
\begin{equation}
E=E_{exc}+E_{Oe}+E_{eb},
\end{equation}
which represents the exchange interaction, Oersted field Zeeman energy, and the exchange bias interaction, respectively. The expressions for the different energy terms are as follows,

\begin{equation}
E_{exc}= A_{pl} d_{pl} \int{d^{2}r \left(\nabla\Phi\right)^{2} }
\end{equation} 

\begin{equation}
E_{Oe}=-\mu_{0} H_{I}\cdot M^{pl}_{s}\int{dV \cos \left(\Phi_{Oe}-\Phi_{\bar{v}}\right)}
\end{equation}

\begin{equation}
E_{eb}=-\mu_{0} H_{bias}\cdot M^{pl}_{s}\cdot d_{pl}\int{d^{2}r   \cos \left(\Phi_{\bar{v}}\right)}
\end{equation}

which can be simplified to give %
\begin{equation}
E_{exc}=4\pi A_{pl} d_{pl} \ln\frac{\| X_{\bar{v}}\| }{r_c},
\end{equation}
\begin{equation}
E_{Oe}=-\mu_{0} M^{pl}_{s} H_{I} \left(\pi d_{pl} r_{nc}X_{\bar{v}}\right)\ln\left[\frac{L}{r_{nc}}\right],
\label{EOe}
\end{equation}
\begin{equation}
E_{eb}= \mu_0\vec{H}_{bias}M^{pl}_{s}d^{pl}\kappa.
\end{equation}
Here, $A_{\text{pl}} d_{\text{pl}}$ is the exchange stiffness (0.4 eV),  $H_{\text{I}} = \left|I \right|/4 \pi r_{\text{nc}}$ represents the Oersted-Amp\`{e}re field, $\Phi_{Oe}$ and $\Phi_{\bar{v}}$ describe the magnetic texture of the Oersted-Amp\`{e}re field and the antivortex respectively, $\mu_{0}M_{\text{s}}^{\text{pl}}$=1.56 T, $d_{\text{pl}}$ represents the thickness of the pinned layer (see Fig. 1), and $\kappa$ represents the effective surface area covered by the vortex-antivortex pair. For a vortex-antivortex separation in the range of 0 to tens of nm, this surface can be approximated by a disk joining the two cores and can be expressed as $\| \vec{X}_{\bar{v}} \|^{2}$.

\begin{figure*}[t]
\includegraphics[width=17cm]{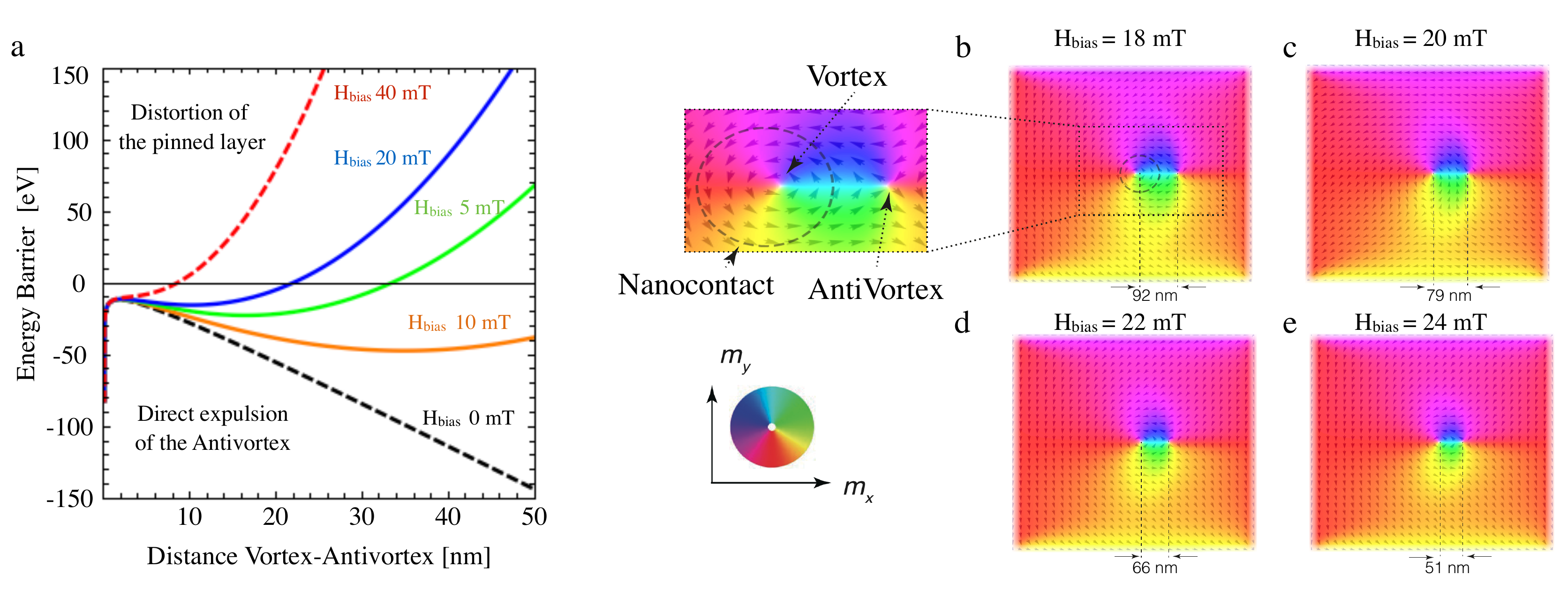}
\caption{\label{energy} (Color online) (a) Topological configuration of the PL for I$_{\text{dc}}=$32 mA. The dashed lines separate the regions without pair, with a pair and with a sole vortex. The ellipses corresponds to the vortex-antivortex separation at equilibrium, with an underestimation factor due to the semi-infinite cylinder approximation and the approximations of the term $\kappa$.Micromagnetic configuration of the PL for I$_{\text{dc}}=$32 mA in the conservative limit for various exchange bias fields (b) 18 mT, (c) 20 mT (d) 22 mT and (e) 24 mT for the exact Oersted field profile.  }
\end{figure*}

Figure~\ref{energy}(a) illustrates how the barrier energy related to pair nucleation depends on the V-AV separation when a dc current of 32 mA (I$_{\text{crit}}$) is applied. 
When the V-AV pair is present, the energy minimum corresponds to a given separation between the cores, which depends on the magnitude of the exchange bias field.  Note that the Oersted-Amp\`{e}re field was calculated considering the NC as an semi-infinite cylinder. This is known ~\cite{APL:Sebastien:2012} to slightly overestimate the Oersted-Amp\`{e}re field, hence the corresponding energy term (eq.~\ref{EOe}); with our material parameter, we will see that this can lead to an underestimation of the separation between the cores of the vortex and the antivortex. 

The model predicts that different magnetic states of the PL can appear depending on the magnitude of the field $\vec{H}_{\text{bias}}$. For $H_{\text{bias}} > $ 40 mT, the energy as function of the V-AV separation distance would exhibit a minimum in the range of the exchange length, $I_{\text{exc}}\approx$ 6 nm. Because this value is much smaller than the spatial extension of the vortex-antivortex pair, it suggests that most likely ground state in a pinned layer with $H_{\text{bias}} > $ 40 mT would be simply a distortion of its magnetization underneath the NC. In the other limit ("unpinned" layer with $H_{\text{bias}} \simeq 0$ mT), the energy minimum would correspond to a V-AV separation distance that falls out of the range showed in Figure~\ref{energy}(a). The preferred state in this case would thus involve the expulsion of the antivortex outside of the nanocontact region with the vortex remaining centered about the nanocontact. For the intermediate cases,  $0 < H_{\text{bias}} < 40$ mT, the exchange bias field leads to an energy minimum at a well-defined separation distance between the vortex-antivortex pair; this distance has a strong dependence on the magnitude of the exchange bias field. We note that under the highest value of the applied current we used in our study (40 mA), the Joule heating leads to a temperature increase of 120 K~\cite{SEbTEMP} above room temperature, which results in a bias field of $H_{\text{bias}}\approx$ 22 mT;  $H_{\text{bias}}$ vanishes at about 600 K in our samples. We conjecture that the separation distance of the vortex-antivortex core remains sufficiently small in our experiment such that spin torques, thermal effects, or both combined can lead to thermally-activated pair annihilation. This provides a mechanism for the intermittence of the LM and UM we observe in the power spectra. 

To shed further light on this scenario, we performed micromagnetics simulations on the nucleation process in the pinned layer using the \textsc{Mumax} code~\cite{Vansteenkiste20112585}. The simulations were performed at zero temperature without the spin-torque terms due to the currents flowing perpendicular to the film plane. The material parameters used were $M_{\text{s}}$=1.260 kA/m, $A=19$ pJ/m, and $\alpha=0.013$. The simulated region was a rectangular volume with dimensions of 1280 nm $\times$ 1280 nm $\times$ 5 nm that was discretized using 512 $\times$ 512 $\times$ 1 finite-difference cells. The spatial distribution of the Oersted-Amp\`{e}re field was computed with full 3D finite-element simulations (COMSOL)~\cite{APL:Sebastien:2012}, which was then included into the micromagnetics simulations.

Results from the micromagnetics simulations are shown in Fig.~\ref{energy}(b)-(e). Our simulations confirm that a stable separation distance between the vortex-antivortex pair is possible for a given value of the applied current and exchange bias field. We notice that the main difference between the analytical model and the simulations is the fact that the antivortex is found to be stabilized outside the nanocontact area while in the analytical model the antivortex remains inside the nanocontact. While there are quantitative differences in the separation distance, the simulation results supports the physical picture that underlies the analytical description.

\subsection{Vortex-antivortex nucleation in the pinned layer in the presence of STT}

The previous simulations neglected spin transfer torques (STT) in the nucleation process of the V-AV pair in the pinned layer. In the present step, we now take into account the STT arising from the intralayer spin currents with a spin polarization constant $P=0.2$; the other system parameters remain unchanged and are still meant to describe a single isolated pinned layer. Results from these simulations are given in Fig. 5.
\begin{figure*}[t]
\includegraphics[width=17cm]{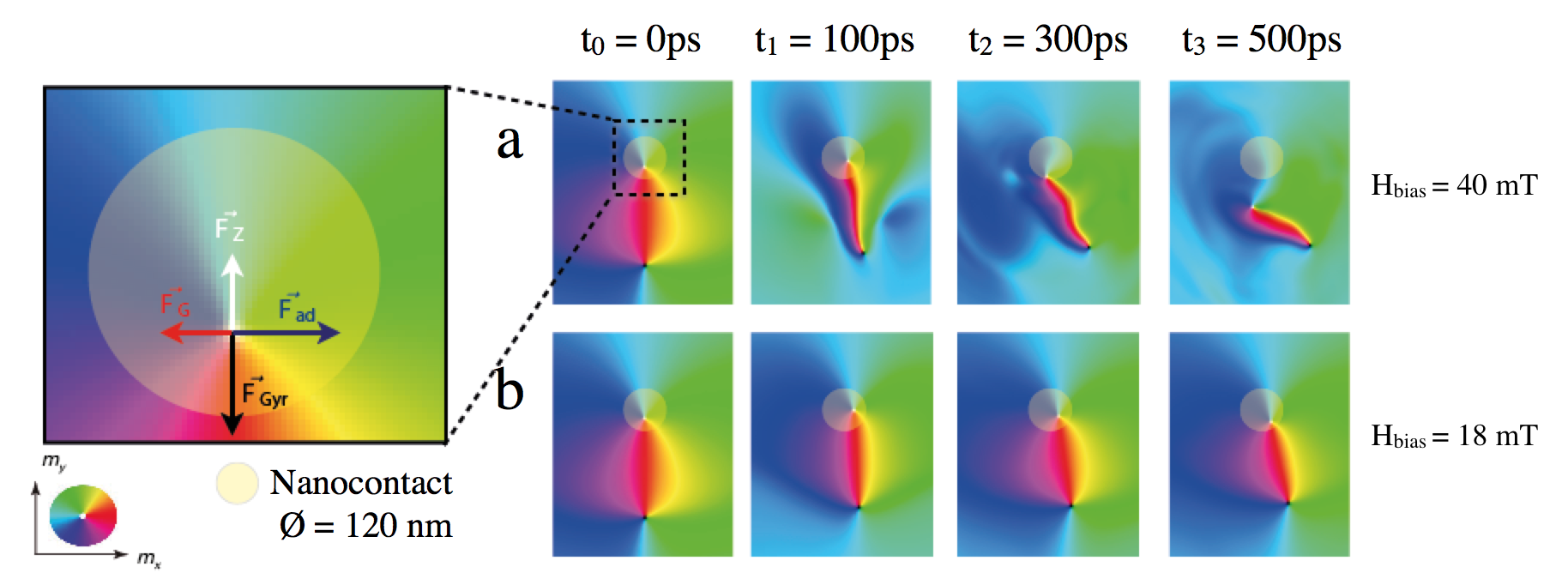}
\caption{\label{STT} (Color online) Micromagnetic simulations showing the time evolution of the vortex-antivortex pair when a constant current $I_\text{dc}$ of 40 mA is applied, for different exchange bias bias field, (a) $\mu_{0}H_{\text{bias}}=$ 40 mT and (b) $\mu_{0} H_{\text{bias}}=$ 18 mT. The intralayer STT is included in this simulation.}
\end{figure*}
As expected, the addition of the STT results in the vortex spiraling out of the edge of the NC. The simulations suggest that the vortex never reaches a stationary orbit because it annihilates with the antivortex first, resulting in a distorted micromagnetic state in the pinned layer. After this annihilation, the process of nucleation can repeat itself. This scenario accounts for the intermittent hopping between the two vortex oscillation modes  (see Fig.~\ref{PSDtemp}). The UM is likely to correspond to the existence of the pair while the LM is likely to correspond to a distortion of the pinned layer (without any out-of-plane component of the magnetization). If the current is considerably larger than $I_{\text{crit}}$, e.g., 50 mA, the antivortex is expelled from the pinned layer and the vortex remaining in the pinned layer starts to perform full rotations around the NC along a stationary orbit. At these very large currents, we would thus be in a situation in which the dynamics inside the pinned layer and inside the free layer are qualitatively similar.

\subsection{Influence of a V-AV pair in the PL onto the oscillating vortex of the free layer}

The analytical theory presented here is based on previous studies \cite{Seb:NatPhys,Kim2012217} in which it was shown that the STT torques from the interlayer spin currents alone cannot drive self-sustained oscillations under zero field. %It was demonstrated that a significant part of the total injected current through the NC flows in the plane of the free magnetic layer. 
Instead, it was shown that this is the in-plane (i.e. intralayer) current that drives the vortex motion. However this was for a static and perfectly in-plane magnetized pinned layer. %To verify it, we reported \cite{APL:Sebastien:2012} that perpendicular current densities (CPP) are restricted below the NC decaying very fast towards the edges. 
%When taking into account the exact in-plane current density (CIP) profile, with its maximum at the NC edges and its slow decay after, the micromagnetic simulations were successfull YYYYYYYY  \cite{Seb:NatPhys}. However, although the understanding has improved with these works, the new experimental facts calls for an extension of this model. 
%Certainly, no previous studies take into account the possible distortion of the pinned layer and its influence on the vortex motion in the free layer. \textbf{JVK: What is the point of this paragraph?}

\begin{figure}[htb]
\includegraphics[width=9cm]{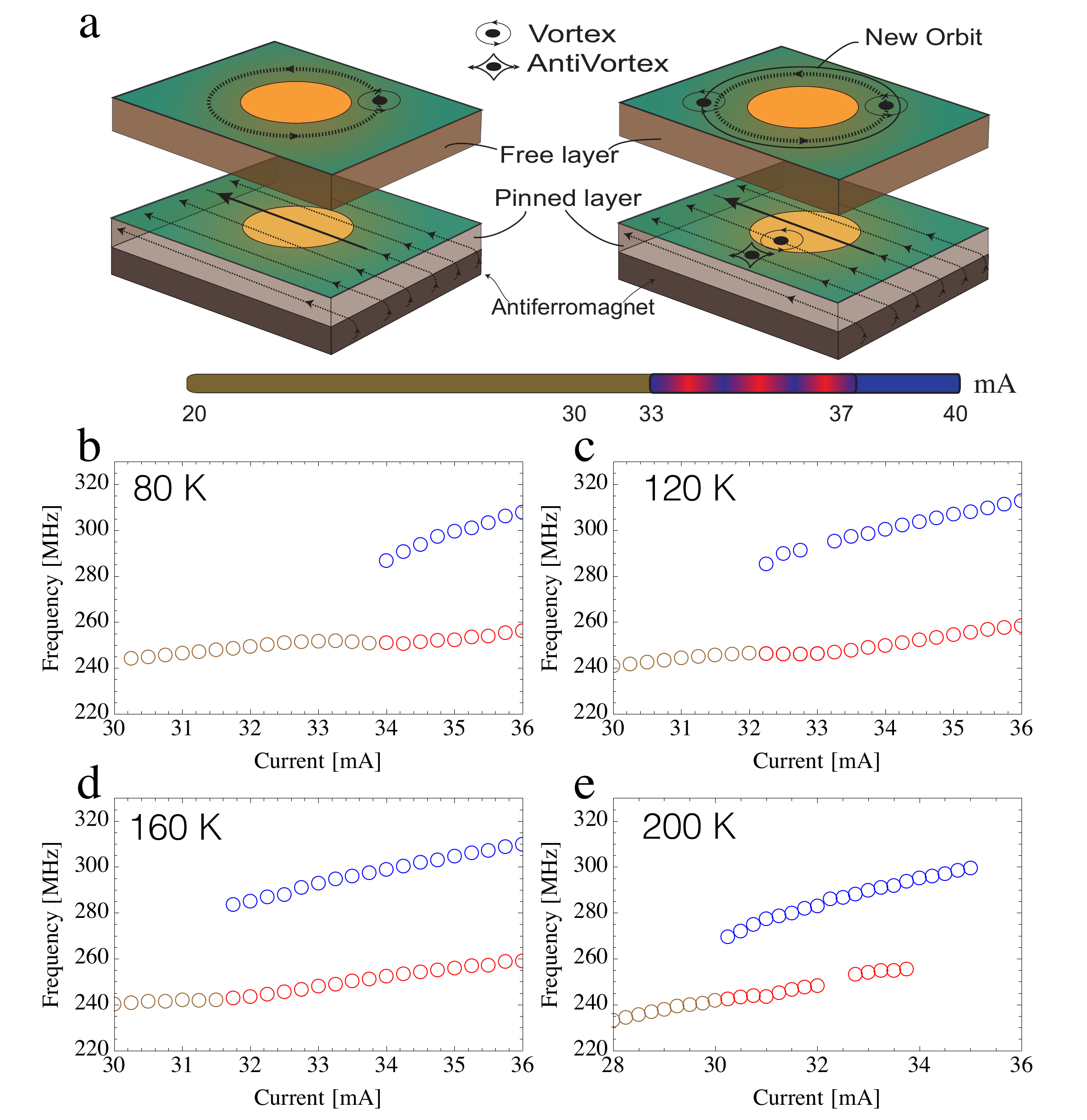}
\caption{\label{result} (Color online) (a) Sketch of the standard vortex mode and its new dynamics when there is a vortex-antivortex pair in the pinned layer. The extended colorbar below the sketch represents the mode at which the vortex is oscillating depending on the applied dc current. Panels (b), (c), (d), (e) show the mode frequency at different temperatures, 80, 120, 160 and 200 K, respectively.}
\end{figure}

Here the presence of the V-AV pair in the PL induces a slight out-of-plane tilt of the spin polarization of the CPP current. To describe the free layer dynamics, we need to take into account this slight out-of-plane tilt of the spin polarization of the CPP current $\vec{P}_{\bot}=p_{\bot}\int dV \text{sin}^{2}\Theta \nabla\Phi$, where ($\Theta,\Phi$) represent the magnetization orientation in polar coordinates. In our specific case, the prefactor of the integral, $p_{\bot}$ is calculated as the ratio between the vortex core radius and the radius of the NC giving a value for $p_{\bot} \approx$ 0.02. The value of $p_{\bot}$ depends on the material parameters in the sense that the vortex core radius scales with the exchange length ($l_{\text{exc}}$) of the magnetic material. Therefore, the larger the saturation magnetization $M_{\text{s}}$, the lower the value of $l_{\text{exc}}$ will be and therefore $p_{\bot}$ is expected to decrease accordingly. The impact of the perpendicular component of spin transfer torque acting over the vortex in the free layer follows Thiele's approach\cite{Thiele}. It includes non conservative torques and it is possible to describe the vortex motion around the NC. To describe the magnetization orientation we use polar coordinates $\Theta(\vec{r})$ and $\Phi(\vec{r})$. $\vec{\textbf{X}}=(X,Y)$ represents the vortex position in the free layer. The equation of motion can be expressed as follows,
\begin{equation}
\label{eq2}
\vec{\textbf{G}}\times
\left[\frac{d\vec{\textbf{X}}}{dt}-\vec{\textbf{u}}(\vec{\textbf{X}})\right]+\hat{D}\left[\alpha\frac{d\vec{\textbf{X}}}{dt}-\beta\vec{\textbf{u}}(\vec{\textbf{X}})\right]+\sigma
I P_{\bot}+\frac{\partial{U_{z}}}{d\vec{\textbf{X}}} =0.
\end{equation}
where $U_{\text{z}}$ represents the Zeeman energy, $\sigma$ represents the spin-torque efficiency, and $\vec{\textbf{u}}(\vec{r})$ the spin-current drift velocity. $\alpha$ and $\beta$ represent the damping constant and the nonadiabatic constant, respectively. By solving Eq. 6, using $R_{0} \exp(i\phi)= X_{0}+iY_{0}$, we find the following coupled differential equations,
\begin{equation}
\label{eq3}
G \partial_{t}R+\alpha D R\partial_{t}\varphi+Gu=\frac{M_{s}\pi r^{2}_{nc}
d_{fl}}{\gamma R}\sigma I P_{\bot} \sin^{2}\Theta_{0},
\end{equation}
\begin{equation}
\label{eq4}
GR \partial_{t}\varphi-\alpha D \partial_{t}R+\beta D u
=\frac{\partial{U_{z}}}{\partial{R}},
\end{equation}
where $d_\text{{fl}}$ and $M_\text{{s}}$, represent the thickness and the saturation magnetization of the free layer, respectively. To determine the radius of the stationary orbit of the vortex we set $\partial_{t}R=0$, which gives
\begin{equation}
\label{eq5}
R_{0}=\frac{\sigma I a^{2}_{nc} G^{2}\left(1+\frac{1}{2}
 p_{\bot} n_{core}\right)}{\alpha
D}\left(\frac{\partial{U_{z}}}{\partial{R}}\right)^{-1}.
\end{equation}
This leads us to classify the FL dynamics into two regimes, depending on the absence ($p_{\bot}$=0) or presence ($p_{\bot} \neq 0$) of a V-AV pair in the PL. The effect of the PL vortex is to increase the radius of the orbit of the FL vortex. Micromagnetic modeling, in agreement with Eq. \ref{eq5}, yields $R_{0}$=110 nm when there is no pair in the PL, and $110 +10 n_{\rm core}$ when cores are present in the PL underneath the NC. The corresponding oscillation frequencies of the FL vortex can be estimated if we assume that the Gilbert damping $\alpha$ and nonadiabatic spin torque parameters $\beta$ have similar magnitudes \cite{beta,beta1}. The frequency tunability, defined as the slope of the frequency versus current relation, then depends on the number of cores ($n_{\text{core}}$) underneath the NC,
\begin{equation}
\label{eq6}
\frac{\partial{\omega_{FL}}}{\partial{I}}=\frac{1}{GR_{0}}\left(\frac{\partial{U_{z}}}{\partial{R}}\right)+\frac{n_{\rm core}}{2}\frac{\alpha
D}{G} \sigma p_{\bot}.
\end{equation}
The agreement between Eq.~\ref{eq6} and the experiments is illustrated in Figure~\ref{result} where we show the dependence of the oscillation frequency versus the applied current. Table I summarizes the theoretical and experimental values for the slope ($\partial\omega_{\text{osc}}/\partial I$) and the frequency jump for different temperatures.

\begin{table}
\caption{\label{tab:table4} Frequency jumps and slopes versus current at different temperatures. LM and UM stand for lower and upper modes. n$_{\text{core}}$ represents the number of vortex or antivortex cores in the pinned layer underneath the nanocontact.}
\begin{ruledtabular}
\newcommand{\mc}[2]{\multicolumn{#1}{c}{#2}}
\begin{tabular}{ccccc}
$T$& $f_{\rm UM} - f_{\rm LM}$ at $I_{\text{crit}} $& $df/dI$ (LM) & $df/dI$ (UM)\\
(K)&(MHz)&(MHz/mA)&(MHz/mA)\\
\hline\\
Theory(0)  &  $\approx$ 36 & $\approx$ 4&$\approx$ 4+2$n_{\text{core}}$\\
\rowcolor{lightgray}
6  & 40 & 4&8.7\\
\rowcolor{lightgray}
80  & 36& 2.8&9\\
\rowcolor{white}
120& 44  & 3.4&5.8\\
\rowcolor{white}
160 & 40 & 3.6 &5.7 \\
\rowcolor{white}
200 & 28 & 4.3 &6
\end{tabular}
\end{ruledtabular}
\end{table}

The validity of the model can be checked by comparing the analytical and experimental slopes. By using the experimental values of $\mu_{0}M_{\text{s}}$, $d_{\text{pl}}$, $\alpha$, and $L$, and by assuming a spin polarization of $P=0.5$ and a radius of the vortex core $r_{\text{core}}$=10 nm, we find a theoretical value for $\partial\omega_{\text{osc}}/\partial I$ =4 MHz/mA in the low frequency regime, which matches with the observed slope experimentally of $\approx$ 3.6 MHz/mA from Eq. (10). The jump in frequency can be calculated theoretically and accounts for a jump of 36 MHz similar to the observed one of $\approx$ 37 MHz. The larger slope in the high frequency regime (upper mode) results from the joint contributions of the Zeeman potential and perpendicular spin torque. This gives a larger slope which again fits well with the experimental one (Table I).

%It can be observed in Table 1 that for low temperatures 6 and 80 K the slope of the upper mode is $\approx$ 30 percent larger than that at higher temperatures. This appears to contradict our previous explanation since the $p_{\bot}$ should be the same for all temperatures (i.e., the vortex profile does not depend on temperature). A plausible explanation would be that the effect of magnetic defects in the nucleation process. Those defects would act as a pinning centers for the antivortex, avoiding the complete expulsion of the antivortex from the NC area. We believe the antivortex (below 80 K) resides together with the vortex underneath the NC region and in this particular case we need also to take into account the profile of the antivortex to compute $p_{\bot}$. Now, $p_{\bot}$ contains the presence of both vortices (vortex-antivortex).
%\textbf{JVK: I'm not sure if this explanation makes any sense. There is no reason why pinning is more or less efficient at different temperatures, unless there is some strong variation in Ms or some other magnetic parameter. TD: I agree that you have to remove this last paragraph. }

\section{Conclusions}
In summary, we have shown that the nucleation of a vortex-antivortex pair in the pinned layer of a spin valve nanocontact oscillator can lead to distinct changes to the oscillatory dynamics of the vortex in the free layer. The pair leads to an additional spin torque term related to currents flowing perpendicular to the film plane, which results in a frequency jump along with a different frequency tunability for the free layer vortex gyration. The temperature dependence of this effect is related to variations in the exchange bias field acting on the pinned layer, which determines the equilibrium separation distance of the vortex-antivortex pair. Intermittent modes in the free layer dynamics are attributed to thermally-activated pair nucleation and annihilation in the pinned layer. Analytical modeling and micromagnetics simulations are shown to give good agreement with experiments.

\begin{acknowledgments}
The authors acknowledge fruitful discussions with G. Hrkac. This work was partially supported by the European Commission under Contract No. MRTN-CT-2008-215368-2 (SEMISPINNET) and the Agence Nationale de la Recherche under Contract No. ANR-09-NANO-006 (VOICE).
\end{acknowledgments}

\bibliography{database}
\end{document}